\documentclass[10pt, conference]{IEEEtran}

\usepackage{multirow}
\usepackage{booktabs}
\usepackage{graphicx}
\usepackage{adjustbox}
\usepackage{listings}
\usepackage[table]{xcolor}
\usepackage{pgfmath}
\colorlet{lightred}    {red!70!white}
\colorlet{lightyellow} {yellow!70!white}
\colorlet{lightgreen}  {green!70!white}
\newcommand{\heatcell}[1]{%
  \pgfmathsetmacro{\val}{#1}%
  \ifdim\val pt<0.5pt
    \pgfmathparse{int(\val/0.5*100)}%
    \edef\heatcolor{%
      \noexpand\cellcolor{lightyellow!\pgfmathresult!lightred}%
    }%
  \else
    \pgfmathparse{int((\val-0.5)/0.5*100)}%
    \edef\heatcolor{%
      \noexpand\cellcolor{lightgreen!\pgfmathresult!lightyellow}%
    }%
  \fi
  \heatcolor \textcolor{black}{#1}%
}

\begin{document}

 \title{LoRA-based Parameter-Efficient LLMs for Continuous Learning in  Edge-based Malware Detection}

\author{
\IEEEauthorblockN{Christian Rondanini\IEEEauthorrefmark{1},
Barbara Carminati\IEEEauthorrefmark{1},
Elena Ferrari\IEEEauthorrefmark{1},
Niccolò Lardo\IEEEauthorrefmark{1}}
\IEEEauthorblockA{\IEEEauthorrefmark{1}DiSTA, University of Insubria, Italy\\
\{christian.rondanini, barbara.carminati, elena.ferrari, nlardo1\}@uninsubria.it}
\and
\IEEEauthorblockN{Ashish Kundu\IEEEauthorrefmark{2}}
\IEEEauthorblockA{\IEEEauthorrefmark{2}Cisco Research, USA\\
\{ashkundu\}@cisco.com}
}

\maketitle

\begin{abstract}
The proliferation of edge devices has created an urgent need for security solutions capable of detecting malware in real time while operating under strict computational and memory constraints. Recently, Large Language Models (LLMs) have demonstrated remarkable capabilities in recognizing complex patterns, yet their deployment on edge devices remains impractical due to their resource demands. However, in edge malware detection, static or centrally retrained models degrade under evolving threats and heterogeneous traffic; locally trained models become siloed and fail to transfer across domains. To overcome these limitations, in this paper, we present a continuous learning architecture for edge-based malware detection that combines local adaptation on each device with global knowledge sharing through parameter-efficient LoRA adapters. Lightweight transformer models (DistilBERT, DistilGPT-2, TinyT5) run on edge nodes and are incrementally fine-tuned on device-specific traffic; only the resulting LoRA modules are aggregated by a lightweight coordinator and redistributed, enabling cross-device generalization without exchanging raw data. We evaluate on two public IoT security datasets, Edge-IIoTset and TON-IoT, under multi-round learning to simulate evolving threats. Compared to isolated fine-tuning, the LoRA-based exchange yields up to 20–25\% accuracy gains when models encounter previously unseen attacks from another domain, while maintaining stable loss and F1 across rounds. LoRA adds less than 1\% to model size (~0.6–1.8 MB), making updates practical for constrained edge hardware.

\end{abstract}


\maketitle

 \section{Introduction}\label{sec:intro}

The rapid evolution of cyber threats and the widespread deployment of edge devices with limited computational resources have made malware detection an increasingly complex challenge.
Traditional centralized security architectures are no longer sufficient, as they introduce latency, reduce scalability, and raise concerns regarding data privacy.
For this reason, the development of local, adaptive, and efficient security mechanisms that operate directly on edge devices has become a pressing priority \cite{xiao2019edge}.

A further complication arises from the dynamic nature of malware itself. 
New attack variants frequently emerge, altering their signatures and behaviors to evade detection mechanisms.
Static models, once trained and deployed, quickly become obsolete, leading to reduced effectiveness in real-world conditions.
This calls for adaptive malware detection systems capable of evolving alongside the threat landscape and integrating new information over time.

Over the past decade, numerous machine learning (ML)-based approaches have been proposed for malware detection, leveraging statistical and neural techniques to identify anomalous network behavior or device-level compromise \cite{ferdous2025survey}.
While these methods have demonstrated strong results in static or centralized settings, they often rely on models trained with fixed datasets or specific traffic distributions, limiting their ability to generalize to unseen attack patterns or heterogeneous environments.

Recent advances in Natural Language Processing (NLP) have demonstrated the ability of Large Language Models (LLMs) to capture complex patterns and adapt to dynamic tasks. 
LLMs are deep neural architectures trained on massive text corpora to learn statistical and semantic representations of language, enabling them to generalize across a wide range of downstream applications such as classification, translation, reasoning, and code generation. 
Their contextual understanding and transfer capabilities make them particularly suitable for tasks such as network traffic analysis, where semantic relationships among features play a crucial role in identifying anomalous or malicious behavior.
Starting from early probabilistic models such as n-grams \cite{brown1992class} and neural architectures like RNNs and LSTMs \cite{mikolov2011strategies,sutskever2014sequence}, the field underwent a paradigm shift with the introduction of the Transformer architecture \cite{vaswani2017attention}. 
The self-attention mechanism enabled transformers to process sequences in parallel and model long-range dependencies, fueling the rise of modern LLMs.
LLMs such as BERT \cite{devlin2018bert}, GPT \cite{gptcore}, and T5 \cite{raffel2020exploring} have since achieved state-of-the-art results in text classification, reasoning, and code analysis. 
These models have demonstrated remarkable capabilities in extracting semantic relationships, identifying complex contextual dependencies, and adapting to a wide variety of downstream tasks.

Applying LLMs to malware detection has attracted increasing research interest, as these models can capture complex patterns in network traffic and accurately identify malicious behaviors \cite{aghaei2022securebert,rahali2021malbert_android,panebianco2025guessing}.
However, most of the existing approaches assume centralized or high-resource environments and are therefore unsuitable for deployment at the edge, where computational capacity, memory, and energy are limited.
Despite challenges in model efficiency and adaptability, deploying LLMs on edge devices offers significant potential for cybersecurity by enabling real-time, privacy-preserving traffic analysis at the local level.

To address these limitations, Small or Lightweight Language Models (SLMs) \cite{xu2024survey} have been investigated as promising candidates, which maintain strong representational capabilities while reducing memory footprint and inference latency through distillation, pruning, or quantization.

Among SLMs, DistilBERT~\cite{sanh2019distilbert}, DistilGPT-2, and T5-Efficient-TINY~\cite{raffel2020exploring} are relevant models that are designed for accuracy and efficiency, which make them suited for resource-constrained and distributed settings.
DistilBERT \cite{sanh2019distilbert} compresses BERT-base by 40\% while retaining about 97\% of its performance, excelling in classification and text comprehension. 
DistilGPT-2, a distilled version of GPT-2, preserves generative capacity with half the parameters, making it suitable for sequential anomaly detection and behavioral modeling. 
T5-Efficient-TINY \cite{raffel2020exploring} extends the encoder–decoder architecture to compact settings, maintaining multi-task versatility while minimizing computational cost.

SLM-based malware detection thus candidates as a promising direction, as it enables the integration of language-inspired contextual reasoning into lightweight, on-device security agents. In our previous work \cite{rondanini2025malware}, we designed an edge-based malware detection architecture that leveraged lightweight and distilled LLMs to balance \textit{feasibility}, \textit{accuracy}, and \textit{adaptability}. 
Feasibility referred to meeting the strict computational, energy, and memory constraints of edge nodes; accuracy addressed the need to maintain reliable detection performance despite model compression; and adaptability captured the capacity to generalize across heterogeneous network flows. 
While that architecture demonstrated the viability of LLM-based malware detection at the edge, it relied on static model updates that limited long-term adaptability.

However, the architecture presented in \cite{rondanini2025malware}  remained essentially static, as local models were trained on specific datasets and their learning processes were confined to their own operational domain. 
This design posed two key limitations.
First, the dynamic and continuously evolving nature of malware requires models capable of incremental adaptation, known as continuous or incremental learning, so that they can assimilate new threat patterns over time without catastrophic forgetting.
Second, the heterogeneous nature of network traffic across different environments causes locally trained models to develop domain-specific knowledge that may not generalize well to other contexts. 
This limitation becomes particularly evident when distinct edge devices, each fine-tuned on different datasets, encounter novel types of attacks unseen during their local training phase.
For instance, when a model fine-tuned on a dataset D1 is evaluated on a different dataset D2, its cross-domain performance remains poor, as each local model learns only from the specific traffic patterns of its training environment.
Our goal is to allow the detection capabilities acquired by a model trained on one network flow to be progressively transferred to others, effectively achieving cross-domain generalization.

Building on these motivations, we propose a continuous learning architecture for edge-based malware detection that combines local adaptability with global knowledge sharing.
The key idea is to enable each node to perform localized fine-tuning on its own traffic while allowing the global system to evolve collaboratively through the exchange of compact, parameter-efficient updates. To coordinate this collaboration, the proposed architecture relies on the presence of a central model, which acts as a collector and aggregator of model updates to be distributed among edge devices.

However, transferring the update by repeatedly fine-tuning full models is computationally prohibitive at the edge. 
To overcome this limitation, parameter-efficient fine-tuning (PEFT) methods have gained attention, enabling selective model updates through modular components. 
Among these, Low-Rank Adaptation (LoRA) \cite{hu2022lora} stands out for its simplicity and effectiveness.
It introduces trainable low-rank matrices within transformer attention layers, freezing the bulk of parameters while ensuring fast and memory-efficient adaptation.
Thus, the learned adaptation is encapsulated in LoRA modules, which are then aggregated and redistributed through a central coordination layer.

In general, to enable collaborative training without data sharing,  decentralized and federated learning (FL/DFL) approaches have been investigated \cite{beltran2023decentralized}.  
However, our architecture pursues a different goal focused on efficient, incremental adaptation through LoRA-based updates. Our design uses a central coordination model to aggregate and redistribute updates, guaranteeing stability and scalability, while DFL necessitates peer-to-peer synchronization and consensus mechanisms (e.g., PBFT). 
For example, \cite{beltran2023decentralized} has observed that DFL experiences irregular convergence and high communication overhead when dealing with non-independent and identically distributed data. 
Our strategy has the goal of achieving easy coordination and effective continuous learning rather than complete decentralization.

To validate the proposed framework, we perform an experimental evaluation on two public datasets, Edge-IIoTset \cite{Ferrag_EdgeIOT} and TON-IoT \cite{moustafa2021new}, which together capture diverse and evolving IoT attack patterns under realistic network conditions.
In our experiments, we simulate the continuous learning process through a series of incremental rounds, each representing a stage in which edge nodes encounter new malware types or updated threat patterns.
At each round, task-specific detection outputs are periodically submitted to the central model, which aggregates, validates, and consolidates the information, updates the shared SLM parameters, extracts them as Low-Rank Adaptation (LoRA) \cite{hu2022lora} modules, and redistributes these updates to all participating devices to enable collective learning and consistent improvement across the network.

\begin{table*}[!htbp]
\centering
\resizebox{\textwidth}{!}{%
\begin{tabular}{@{}lll@{}}
\toprule
\textbf{Category} & \textbf{Principle} & \textbf{Representative Techniques} \\ \midrule
\multicolumn{1}{|l|}{Parameter-Efficient Fine-Tuning} &
  \multicolumn{1}{l|}{\begin{tabular}[c]{@{}l@{}}Update a limited set of parameters \\ while keeping the backbone frozen\end{tabular}} &
  \multicolumn{1}{l|}{\begin{tabular}[c]{@{}l@{}}LoRA, QLoRA, Adapters, \\ Prefix/Prompt-Tuning, BitFit\end{tabular}} \\ \midrule
\multicolumn{1}{|l|}{Memory-Based Replay} &
  \multicolumn{1}{l|}{\begin{tabular}[c]{@{}l@{}}Preserve past knowledge \\ by reusing stored or generated data\end{tabular}} &
  \multicolumn{1}{l|}{\begin{tabular}[c]{@{}l@{}}Experience Replay, Generative Replay, \\ Embedding Memory\end{tabular}} \\ \midrule
\multicolumn{1}{|l|}{Regularization-Based} &
  \multicolumn{1}{l|}{\begin{tabular}[c]{@{}l@{}}Constrain parameter updates \\ to retain prior knowledge\end{tabular}} &
  \multicolumn{1}{l|}{\begin{tabular}[c]{@{}l@{}}Elastic Weight Consolidation (EWC), \\ Learning without Forgetting (LwF), \\ Synaptic Intelligence\end{tabular}} \\ \midrule
\multicolumn{1}{|l|}{Modular Architectures} &
  \multicolumn{1}{l|}{\begin{tabular}[c]{@{}l@{}}Introduce specialized or expandable \\ components for new tasks\end{tabular}} &
  \multicolumn{1}{l|}{Mixture-of-Experts (MoE), Progressive Networks} \\ \midrule
\multicolumn{1}{|l|}{Retrieval-Augmented Learning} &
  \multicolumn{1}{l|}{\begin{tabular}[c]{@{}l@{}}Offload knowledge to external \\ retrieval mechanisms or tools\end{tabular}} &
  \multicolumn{1}{l|}{\begin{tabular}[c]{@{}l@{}}Retrieval-Augmented Generation (RAG), \\ Vector Databases, Tool-Augmented Models\end{tabular}} \\ \bottomrule
\end{tabular}%
}
\vspace{+0.1em}
\caption{Approaches to Continuous Learning in Large Language Models}
\label{tab:cl_methods}
\end{table*}

The rest of this paper is organized as follows. 
Section \ref{sec:continuos} discusses continuous learning principles and their relevance for malware detection. 
Section \ref{sec:challenges+architecture} presents the proposed architecture.
Section \ref{sec:applications} reviews related work.
Section \ref{sec:expresults} reports and analyzes the experimental results. 
Finally, Section \ref{sec:conclusions} concludes the paper and outlines directions for future research.

\section{Continuous Learning}\label{sec:continuos}

Continuous learning, also known as lifelong or incremental learning, has become an essential research direction for enabling large language models (LLMs) to adapt to evolving data distributions and novel tasks without retraining from scratch. 
Unlike traditional fine-tuning, which operates under the assumption of a static dataset and training objective, continuous learning emphasizes integrating new knowledge over time while minimizing catastrophic forgetting, ensuring scalability, and preserving generalization. 
Several strategies have been proposed in the literature, which classify them into parameter-efficient fine-tuning, memory-based replay, regularization-based methods, modular architectures, and retrieval-augmented learning \cite{zheng2025towards}.
Table~\ref{tab:cl_methods} summarizes these categories and their representative techniques.

As introduced in Section \ref{sec:intro}, we aim at designing a framework where each node performs local detection on device-specific data while contributing its learned knowledge to a shared, evolving global model.
As edge environments impose strict trade-offs between adaptability, efficiency, and resource consumption, choosing the most suitable continuous learning approach becomes crucial to achieving such a setting.
Accordingly, in the following, we provide a detailed discussion of each major category of continuous learning strategies and their representative methods, emphasizing their applicability within edge-based malware detection scenarios.

\subsection{Continuous learning in LLM}

Parameter-efficient fine-tuning (PEFT) \cite{han2024parameter} aims to adapt LLMs by updating only a small subset of parameters while keeping most of the backbone frozen.
This approach's key advantage is reducing the computational cost, accelerating training, and allowing for modular adaptation across tasks. 
Among the most relevant PEFT methods is Low-Rank Adaptation (LoRA) \cite{hu2022lora}, which introduces rank-decomposed trainable matrices into the attention mechanism. 
LoRA significantly reduces memory and parameter requirements by approximating weight updates through low-rank factorization.
An extension of this approach, QLoRA \cite{dettmers2023qlora}, further compresses models by quantizing their weights, enabling efficient adaptation even on resource-constrained hardware. 

\begin{table*}[!htbp]
\centering
\begin{tabular}{p{2cm} p{2.1cm} p{4.5cm} p{7.4cm}}
\toprule
\textbf{Approach \newline Category} & \textbf{ Method} & \textbf{Strengths} & \textbf{Limitations in Our Context} \\
\midrule
\multirow{4}{2.5cm}{\textbf{Parameter-\newline Efficient Fine-Tuning \newline (PEFT)}} & 
    LoRA / QLoRA & 
    Minimal computational and memory cost \newline 
    Modular per-task adaptation \newline
    Seamless integration with transformers (e.g., DistilBERT, TinyT5) & 
    Requires managing multiple adapters, which may increase the memory footprint on edge devices \newline 
    Prediction aggregation across adapters can be complex and may introduce latency for real-time detection \\
\cmidrule(lr){2-4}
& Adapters & 
    Modular and task-specific adaptation \newline 
    No full model retraining required & 
    Extra parameters per task must be carefully managed across tasks \newline 
    May underperform on malware classification tasks \newline 
    Storage overhead on highly constrained devices \\
\cmidrule(lr){2-4}
& Prefix / \newline Prompt Tuning & 
    Extremely lightweight tuning \newline 
    Minimal parameter storage & 
    Less effective for classification tasks, which degrades accuracy in malware detection \newline 
    Sensitive to evolving malware patterns \\
\cmidrule(lr){2-4}
& BitFit & 
    Trains only bias terms \newline 
    Minimal parameter cost & 
    Limited expressive power \newline
    Fail to capture complex or evolving malware behaviors \\
\midrule
\multirow{2}{2.5cm}{\textbf{Memory-Based Replay}} & 
    Standard Replay & 
    Prevents forgetting \newline 
    Simple conceptually & 
    Requires storing past data, which is infeasible on edge devices for malware traffic or binaries \newline 
    Raises privacy concerns and conflicts with compliance requirements \\
\cmidrule(lr){2-4}
& Generative Replay & 
    Reduces storage need \newline 
    Avoids direct storage of sensitive data & 
    Generated samples may be imperfect \newline 
    SLM may fail to accurately reproduce past malware patterns \newline 
    High computational cost and generation overhead unsuitable for edge deployment \\
\midrule
\multirow{3}{2.5cm}{\textbf{Regularization-Based}} & 
    Elastic Weight \newline Consolidation (EWC) & 
    No old data storage required \newline 
    Preserves previous knowledge & 
    Sensitive to hyperparameters, making tuning infeasible on edge hardware \newline 
    Struggles as task number grows \newline
    Does not scale with continuously evolving malware classes \\
\cmidrule(lr){2-4}
& Learning without \newline Forgetting (LwF) & 
    Avoids data storage \newline 
    Simple neural architectures integration& 
    Underperforms on highly divergent tasks \newline 
    Limited capacity in SLMs to preserve earlier detection patterns \\
\cmidrule(lr){2-4}
& Synaptic Intelligence & 
    Balances stability and plasticity \newline 
    No data replay required & 
    Sensitive to tuning with additional computational overhead that exceeds resource limits on edge devices \newline 
    Limited performance on novel attack types \\
\midrule
\multirow{2}{2.5cm}{\textbf{Modular Architectures}} & 
    Mixture-of-Experts (MoE) & 
    Task isolation \newline 
    Enables transfer across tasks & 
    Model size grows with number of tasks, making it infeasible for edge deployment \newline 
    High resource cost, too heavy for distilled models \\
\cmidrule(lr){2-4}
& Progressive Networks & 
    Preserves old task knowledge \newline 
    Supports incremental expansion & 
    Model complexity increases rapidly with high resource usage that exceeds edge capabilities \newline 
    Not suitable for real-time malware detection \\
\midrule
\textbf{Retrieval-Augmented Learning} & 
    Vector Database / \newline Tool-Augmented (RAG) & 
    Dynamic access to external knowledge \newline 
    No retraining for new information & 
    Relies on retrieval rather than parameter updates, making it less suited for classification tasks \newline 
    Malware detection requires internal feature learning, not retrieval \newline 
    Inefficient for adapting to new attack patterns \\
\bottomrule
\end{tabular}
\vspace{+0.1em}
\caption{Comparison of continuous learning approaches in the context of edge-based malware detection with small language models.}
\label{tab:comparison}
\end{table*}

Other important PEFT methods include adapters, which are lightweight neural modules inserted between existing layers of the model. 
After training, the base model stays the same, but these adapters can be switched or stacked for task-specific adaptation.
Moreover, prefix tuning \cite{li2021prefix} and prompt tuning \cite{Lester2021ThePO} add more learnable tokens at the input or attention level, guiding the model's actions without changing its internal parameters. 

Finally, BitFit \cite{zaken2021bitfit} represents an extreme case of parameter efficiency, where only bias terms (i.e., the additive constants in neural network layers)  are fine-tuned, offering a simple and effective mechanism for task adaptation. 

Another technique is Memory-based replay, which preserves past knowledge by maintaining a memory buffer of examples from earlier tasks and replaying them alongside new data during training.
The simplest approach is experience replay \cite{rolnick2019experience}. It saves representative samples from previously viewed data and periodically replays them alongside new data to prevent forgetting.
However, careful memory management is required to balance storage constraints with task coverage.  
Generative replay \cite{shin2017continual} is an alternative, in which a generative model generates pseudo-samples that replicate data from earlier tasks. 
Although explicit data storage is no longer necessary, this presents the problem of preserving generative fidelity over time. 
An alternative approach called embedding memory \cite{lopez2017gradient} focuses on storing condensed representations rather than raw data, making it possible to retrieve historical data while adapting efficiently. 
The key idea across these techniques is to stabilize learning in incremental scenarios by anchoring the model to prior distributions.

Regularization-based methods constrain parameter updates during training to preserve knowledge from earlier tasks. 
One of the most extensively researched methods is Elastic Weight Consolidation (EWC) \cite{xiang2023language}, which 
penalizes significant deviations during new learning and uses the Fisher information matrix to estimate the importance of each parameter for previous tasks (i.e., a way to measure how much changing a parameter would affect the model's predictions).
In this way, parameters critical to past tasks remain stable while allowing plasticity in less important regions.  
Another approach, Learning without Forgetting (LwF) \cite{li2017learning}, employs knowledge distillation by preserving the output distribution of a model trained on previous tasks. 
To preserve previous knowledge without needing access to historical data, the model is jointly optimized on new objectives while matching the responses of its predecessor.
Similarly, Synaptic Intelligence \cite{zenke2017continual} measures the contribution of parameters during training and penalizes disruptive updates, striking a balance between stability and plasticity. 

Modular approaches to continuous learning emphasize structural adaptation, where models expand or reconfigure their architectures to accommodate new tasks. 
A notable example is the Mixture-of-Experts (MoE) \cite{yu2024boosting} framework, which activates a subset of specialized experts  (independent neural modules) for each input. 
By routing tasks to appropriate experts, MoE improves scalability and efficiency while reducing interference between tasks.  
Another modular approach is represented by progressive networks, which sequentially add new modules for new tasks while freezing previously learned modules. 
Connections between new and old modules allow for the transfer of useful representations while preventing forgetting of earlier knowledge. 
Although modularity improves robustness and scalability, it also poses challenges with resource efficiency and model growth, necessitating careful architectural design.

Retrieval-augmented learning represents a complementary paradigm in which adaptation occurs not by updating parameters but by leveraging external sources of knowledge. 
In Retrieval-Augmented Generation (RAG), the model retrieves relevant documents or facts from a database at inference time, conditioning its output on external information without requiring retraining.
This approach enables rapid adaptation to new information and domains.  
Closely related methods include \textit{vector databases}, which provide dense retrieval mechanisms for storing and accessing embeddings of prior knowledge. 
By aligning queries with stored representations, models can recall relevant past information dynamically. 
Furthermore, tool-augmented models integrate external systems, such as APIs or symbolic reasoners, to extend reasoning and task-specific abilities without modifying the core model.
By shifting adaptation from weights to retrieval, these methods reduce the risk of forgetting and support continual learning in dynamic environments.

The choice among these methods ultimately depends on the deployment context: the available computational resources, the severity of the distribution shift, and the acceptable trade-off between plasticity and stability.

\subsection{LoRA for edge continual malware Detection}\label{subsec:why_lora}
 
Selecting an appropriate strategy for continuous learning in edge environments requires balancing efficiency, modularity, and adaptability to evolving attack classes. 
Traditional full fine-tuning approaches are computationally expensive and prone to catastrophic forgetting, making them unsuitable for iterative adaptation on constrained hardware. 

A comparative overview of existing continuous learning paradigms is provided in Table~\ref{tab:comparison}. 
While effective at preventing forgetting, memory-based replay techniques are incompatible with edge-based malware detection. 
Standard replay requires storing past malware samples, which is infeasible on resource-constrained devices and raises significant privacy and compliance concerns. 
Generative replay attempts to mitigate storage requirements by synthesizing past data, but the computational overhead of generation and the risk of imperfect sample reproduction make it unsuitable for real-time detection scenarios, particularly when using small language models that may fail to capture complex malware patterns accurately.

Regularization-based approaches such as Elastic Weight Consolidation (EWC), Learning without Forgetting (LwF), and Synaptic Intelligence avoid data storage by constraining weight updates to preserve previous knowledge. 
However, these methods introduce critical limitations in our context. 
EWC requires careful hyperparameter tuning to balance stability and plasticity, a process that is computationally infeasible on edge devices and becomes increasingly difficult as the number of malware classes grows. 
LwF relies on knowledge distillation from previous task predictions, but the limited capacity of small language models makes it challenging to preserve earlier detection patterns when faced with highly divergent attack types. 
Synaptic Intelligence, while theoretically balancing parameter importance, incurs additional computational overhead that exceeds the resource constraints of edge deployment and demonstrates limited effectiveness on novel, previously unseen attack patterns.

Modular architectures like Mixture-of-Experts (MoE) and Progressive Networks offer strong task isolation and knowledge preservation, but their model complexity grows linearly or super-linearly with the number of tasks. 
This architectural expansion rapidly exceeds both the memory capacity and computational budget available on edge devices, making them impractical for scenarios requiring continuous adaptation to emerging malware families. 

Similarly, Retrieval-Augmented Generation (RAG) techniques are not in line with malware classification requirements, even though they are effective for knowledge-intensive tasks. 
RAG systems rely on retrieving relevant information from external databases rather than learning discriminative features within model parameters, making them inefficient for adapting to new attack patterns and unsuitable for tasks requiring rapid, local inference without network access.

Parameter-Efficient Fine-Tuning (PEFT) methods address these issues by updating only a fraction of model parameters while keeping the backbone frozen, ensuring minimal memory and compute requirements.
Unlike other continuous learning paradigms, PEFT approaches inherently scale more efficiently, as the selective update of a small parameter subset avoids the exponential growth in memory and computation typically associated with replay buffers, regularization terms, or modular expansions.
Among PEFT strategies, \textbf{Low-Rank Adaptation (LoRA)} provides the most balanced trade-off between adaptability and efficiency for our specific use case. 
By injecting low-rank trainable matrices within transformer attention layers, LoRA enables modular, per-task updates with minimal parameter overhead, typically adding only 0.1-1\% of the original model parameters per adapter. 
LoRA works in parallel with the model's current weights, maintaining the model's representational capacity while retaining computational efficiency, in contrast to standard adapters that add more sequential layers and might perform poorly on classification tasks. 
Compared to prefix or prompt tuning methods, which manipulate input representations and are inherently less effective for discriminative classification tasks, LoRA directly modifies attention mechanisms where malware-distinguishing features are learned. 
Furthermore, LoRA adapters can be independently trained, stored, and dynamically loaded per malware class or detection round, enabling true modularity without the aggregation complexity or latency issues associated with managing multiple adapter types. 
\textit{This modular design is critical for scenarios} where different edge nodes may encounter malware families in varying orders, allowing each device to maintain only the adapters relevant to its local threat landscape while supporting efficient knowledge sharing through adapter distribution across the network.

\begin{figure*}[!htbp]
\centering
 \includegraphics[width=0.75\textwidth, keepaspectratio]{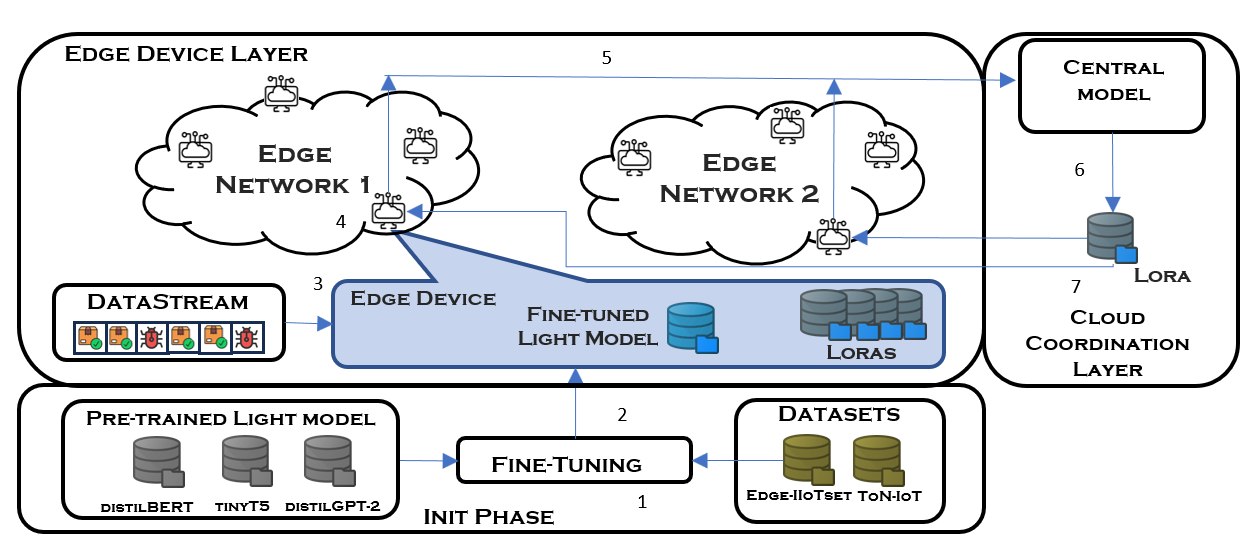}
 \caption{Proposed architecture for LoRA-based parameter-efficient continual learning in edge malware detection.}\label{fig:EdgeLayerArchitecture}
 \end{figure*}

The quantized variant, QLoRA, further reduces memory requirements by performing adapter training on quantized base models, making it particularly suitable for deployment on edge devices with limited RAM. 
This combination of minimal computational cost, modular adaptability, seamless integration with transformer architectures, and the ability to mitigate catastrophic forgetting without data retention makes LoRA the optimal choice for our edge-based malware detection framework.

\section{LoRA-Based  Architecture for  Edge-based  Continous Malware Detection}\label{sec:challenges+architecture}

Deploying LLMs for malware detection at the edge enables real-time threat analysis close to data sources, but despite our previous work demonstrating the feasibility of using lightweight, distilled models to balance efficiency and accuracy, their static nature still limits long-term adaptability in dynamic environments.

To overcome this limitation, we extend that foundation by introducing a continuous learning paradigm. 

While a central coordination layer gathers and disseminates knowledge throughout the network, each edge node in this new configuration carries out local detection on its own data streams, capturing the distinct malware behaviors seen in their operational environments. 
Thus, rather than deploying static models that degrade over time, the proposed framework enables incremental updates through parameter-efficient fine-tuning.

However, enabling continuous learning at the edge also introduces the challenge of knowledge isolation: locally trained models may adapt effectively to their immediate environments but remain unaware of threats detected elsewhere in the network.
A central coordinating model is needed to aggregate, verify, and redistribute the locally learned knowledge in order to guarantee that the system evolves collectively rather than breaking up into separate learners.
This coordination layer acts as a bridge among edge nodes, allowing updates learned in one context to propagate to others without exchanging raw or sensitive traffic data.

As envisioned in our earlier work \cite{rondanini2024large}, a promising solution is to adopt a semi-decentralized design, where the central model serves as a lightweight orchestrator rather than a monolithic controller.
In this configuration, edge devices retain full autonomy for local adaptation, while the central layer consolidates LoRA-based parameter updates and selectively disseminates them across the network.


The proposed architecture, illustrated in Figure~\ref{fig:EdgeLayerArchitecture}, extends our previous centralized design \cite{rondanini2025malware,rondanini2024large} with a continuous learning system composed of two layers: the \textit{Edge Device Layer} and the \textit{Cloud Coordination Layer}.  

The detection process begins at the device layer, with the initialization of lightweight pre-trained language models such as DistilBERT, DistilGPT-2, or TinyT5, which are selected for their balance between representational power and computational efficiency (step 1, Figure~\ref{fig:EdgeLayerArchitecture}). 
Each model is then fine-tuned on a dataset, such as Edge-IIoTset or TON-IoT, to capture domain-relevant features and adapt to the distinct behavioral patterns of each operational environment (step 2, Figure~\ref{fig:EdgeLayerArchitecture}). 
Following the dataset-specific fine-tuning, the optimized lightweight models are deployed across multiple edge devices in different networks (step 3, Figure~\ref{fig:EdgeLayerArchitecture}). 
Each device performs malware detection on a data stream (e.g., network flow).
The resulting task-specific detection performed by the devices is periodically submitted to the \textit{central model} in the cloud layer (step 4, Figure~\ref{fig:EdgeLayerArchitecture}). 

The central model periodically aggregates, validates, and consolidates the received information, updating the SML model parameters and extracting them as Low-Rank Adaptation (LoRA) modules (step 5, Figure~\ref{fig:EdgeLayerArchitecture}).
These LoRA capture the essential parameter shifts required to adapt the frozen backbone to local threats while remaining lightweight, typically only a few megabytes compared to hundreds for full model updates.
Then, the updated LoRA modules are distributed back to the edge networks, where each device can then incorporate these refined adapters into its local backbone, effectively benefiting from the experiences and knowledge gained by other devices and networks without exchanging any raw or sensitive data (step 7, Figure~\ref{fig:EdgeLayerArchitecture}).

To illustrate the practical benefits of our approach, let us consider a realistic deployment scenario with two edge devices, both initially fine-tuned on DistilBERT but trained on different datasets: \textbf{Device1} is trained on the \textit{Edge-IIoTset}, while \textbf{Device2} is trained on the \textit{TON-IoT} dataset.
This setup, depicted in Figure \ref{fig:roundimprovement}, reflects a common challenge in distributed IoT security, where individual devices have limited exposure to the full spectrum of threat patterns due to localized data collection. 
At round 0, each device can only detect malware present in its own training dataset, creating blind spots in its respective detection capabilities. 
However, after just completing round 1, both devices gain the ability to detect threats from both datasets. 
By design, our architecture ensures continuous improvement in detection capability with each subsequent round, progressively expanding the threat coverage across all participating devices while maintaining the efficiency benefits of LoRA-based parameter updates.

Through this architecture, learning becomes both iterative and bidirectional: the edge layer continuously adapts to its immediate environment, while the central layer synchronizes and disseminates collective improvements across the system.

\section{Experimental Results}\label{sec:expresults}

\begin{figure}[!htbp]
\centering
 \includegraphics[width=0.83\columnwidth, keepaspectratio]{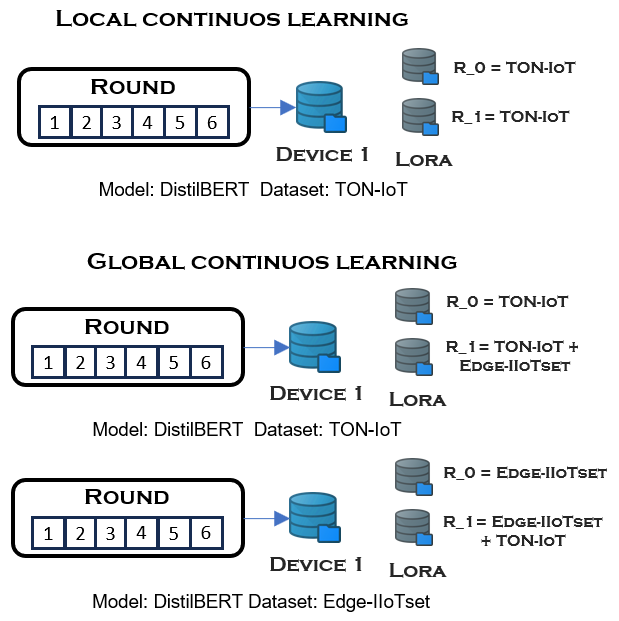}
 \caption{Detection capability example for local and  continuous learning approaches}\label{fig:roundimprovement}
 \end{figure}

In this section, we present the experimental evaluation conducted to validate the effectiveness of the proposed continuous learning architecture.
In designing the evaluation, we intentionally exclude the evaluation of runtime overhead and energy consumption, as those aspects were already extensively analyzed in our prior work \cite{rondanini2025malware} under comparable conditions, where we demonstrated that lightweight LLMs can operate efficiently on edge devices such as Raspberry Pi and Jetson Nano with negligible latency overhead. In \cite{rondanini2025malware}, results showed that lightweight transformer architectures such as DistilBERT and TinyBERT could be executed efficiently, achieving inference latencies below 200 ms and CPU utilization under 45\%, while maintaining F1-scores up to 0.999 in domain-specific detection tasks. Given these results, this work focuses exclusively on detection performance metrics. Specifically, as depicted in Figure \ref{fig:roundimprovement}, our experiments aim to validate two key aspects:
\begin{enumerate}
    \item {\em Local continuous learning}: the capability of each local model to perform continuous learning, that is, to incrementally acquire knowledge from its own operational environment and improve detection performance after encountering new malware types.
    \item {\em Global continuous learning}: the effectiveness of global continuous learning, where knowledge learned by one model is successfully transferred to others through LoRA-based updates, enabling devices to detect threats that were never directly observed in their local datasets.
\end{enumerate}

\subsection{Experiment Configuration}

In this section, we describe the approach used to compute the effectiveness of the proposed malware detection framework (e.g., accuracy and F1 score), as well as the adopted datasets.

\subsubsection{Classification configuration}
The classification procedure adopted in this study follows the same operational logic as in our previous work \cite{rondanini2025malware}.
Each data instance, consisting of feature–value pairs (e.g., \texttt{tcp.dstport:1883}, \texttt{mqtt.conflag.cleansess:1}), is tokenized and provided as textual input to the transformer-based model (e.g., DistilBERT, DistilGPT-2, or TinyT5). 
The model then encodes the contextual dependencies among these features through its attention mechanisms and predicts the corresponding class label (benign or the attack type).

\textbf{Models initialization.}
Experiments were conducted starting from multiple pre-trained language models.
Each pre-trained model was retrieved for the Pytorch framework from the official repository on Huggingface\footnote{https://huggingface.co/}, which can be considered a de facto standard for the open-source code of NLP and machine learning models.
The models considered in our experiments include: 
DistilGPT2\footnote{https://huggingface.co/distilbert/distilgpt2}, 
DistilBERT\footnote{https://huggingface.co/distilbert/distilbert-base-uncased},
and TinyT5\footnote{https://huggingface.co/t5-efficient-tiny}.
Since these pre-trained models are headless (i.e., they do not include a task-specific classification layer), an initial fine-tuning step is required to attach and train a classification head on labeled data. 
Accordingly, each model is first fine-tuned on a distinct starting dataset.
Model initialization utilized pre-trained models from the Hugging Face repository, configuring classification heads according to the number of distinct malware classes (e.g., $GPT2ForSequenceClassification$ for distillGPT-2).
Models and tokenizers are then loaded onto the target computing devices.
Training configuration employed the $AdamW$ optimizer with a learning rate of $2 \times 10^{-5}$ and $CrossEntropyLoss$ as the loss function, combining LogSoftmax with negative log-likelihood for multiclass classification. 
The training procedure is executed on at least four epochs, with each epoch comprising forward propagation, loss computation, back-propagation, and optimization steps.
Validation occurred after each epoch to monitor performance metrics.

\begin{table}[!htbp]
\centering
\resizebox{\columnwidth}{!}{%
\begin{tabular}{@{}|c|c|c|c|}
\midrule
\multicolumn{2}{|c|}{\textbf{Train set}} & 
\multicolumn{2}{|c|}{\textbf{Test set}} \\ \midrule
\textbf{Target} & \textbf{Num samples} &
\textbf{Target} & \textbf{Num samples} \\ \midrule

Normal        & 145 & Normal        & 105 \\ \midrule
DDoS\_UDP     & 155 & DDoS\_UDP     & 95  \\ \midrule
Password      & 146 & Password      & 104 \\ \midrule
XSS           & 142 & XSS           & 108 \\ \midrule
Backdoor      & 152 & Backdoor      & 98  \\ \midrule
SQL injection & 136 & SQL injection & 114 \\ \midrule
Fingerprinting& 150 & Fingerprinting& 100 \\ \midrule
MITM          & 158 & MITM          & 92  \\ \midrule
Port Scanning & 156 & Port Scanning & 94  \\ \midrule

Uploading & 155 & Uploading & 95  \\ \midrule
DDoS\_TCP & 142 & DDoS\_TCP & 108  \\ \midrule
DDoS\_ICMP & 155 & DDoS\_ICMP & 95  \\ \midrule
DDoS\_HTTP & 153 & DDoS\_HTTP & 97  \\ \midrule
Ransomware & 156 & Ransomware & 94  \\ \midrule
Vulnerability\_scanner & 149 & Vulnerability\_scanner & 101  \\ \midrule

\end{tabular}%
}
\caption{Combined  sample distribution for train and test of the datasets }
\label{tab:malwareDistribution}
\end{table}

\textbf{Single-round classification.} In the single-round configuration, each model is evaluated independently on a specific malware classification task without any incremental learning.
This setup serves as a baseline scenario, reflecting a traditional static training–testing pipeline where the model is fine-tuned once on a dataset and evaluated on the corresponding test split.
During this process, the model learns to associate network traffic samples with their respective attack categories.


\begin{table}[!htbp]
\centering
\resizebox{0.80\columnwidth}{!}{%
\begin{tabular}{|c|c|}
\midrule
\textbf{Round} & \textbf{New classes} \\ \midrule
0 & Normal, DDoS\_UDP, Password \\ \midrule
1 & XSS, Backdoor \\ \midrule
2 & SQL\_injection, Fingerprinting \\ \midrule
3 & MITM, Port\_Scanning \\ \midrule
4 & Uploading, DDoS\_TCP \\ \midrule
5 & DDoS\_ICMP, DDoS\_HTTP \\ \midrule
6 & Ransomware, Vulnerability\_scanner\\ \midrule
\end{tabular}
}
\caption{Example of attacks added in each round}
\label{tab:Malware_for_Rounds}
\end{table}

The model's performance is evaluated using the following metrics: accuracy, precision, recall, and F1-score.
In their definitions, TP (true positives) denotes data items  (i.e., network flow) correctly labeled as positive, that is, with malware; FN (false negatives) denotes the data items incorrectly labeled as negative when they are actually positive;  FP (false positives) denotes those labeled as positive when they are actually negative; whereas TN (true negatives) denotes the number of those correctly labeled as negative, i.e., without malware.
To verify the absence of overfitting, we also applied 5-fold cross-validation, obtaining results consistent with those reported in our previous study \cite{rondanini2025malware}.

\textbf{Multi-round classification.} To simulate continuous learning, the training process is extended across multiple rounds.
We considered all malware families from both datasets, incrementally introducing at least two families in each round, until covering all the malware.
We conducted the test multiple times, varying the order in which malware families were introduced across rounds, with an example configuration depicted in Table \ref{tab:Malware_for_Rounds}.
By design, for every two newly introduced attack types, a new LoRA adapter is generated.
During the evaluation phase, the test set is processed sequentially through all rounds: for each round, the corresponding LoRA adapter and its classification head are dynamically loaded.
Each model outputs logits corresponding to the attack classes learned up to that round. 
These scores are collected into a unified structure representing all classes known up to that point. 
The final prediction for each instance corresponds to the class with the highest normalized logit value.

\begin{table*}[!htbp]
\centering
\resizebox{0.90\textwidth}{!}{%
\begin{tabular}{@{}|c|c|c|c|c|c|c|c|c|c|c|c|@{}}
\midrule
\multicolumn{2}{|c|}{} & \multicolumn{5}{c|}{\textbf{Round 0}} & \multicolumn{5}{c|}{\textbf{Round 1}} \\ \midrule
\textbf{Model} & \textbf{Epoch} & \textbf{Loss} & \textbf{Accuracy} & \textbf{Precision} & \textbf{Recall} & \textbf{F1-score} 
& \textbf{Loss} & \textbf{Accuracy} & \textbf{Precision} & \textbf{Recall} & \textbf{F1-score} \\ \midrule

\multirow{5}{*}{\textbf{DistilBERT}} 
& 1  & 1.012 & \heatcell{0.462} & 0.490 & 0.462 & \heatcell{0.455} & 0.934 & \heatcell{0.478} & 0.485 & 0.478 & \heatcell{0.471} \\ \cmidrule{2-12}
& 2  & 0.842 & \heatcell{0.582} & 0.587 & 0.582 & \heatcell{0.579} & 0.768 & \heatcell{0.592} & 0.596 & 0.592 & \heatcell{0.590} \\ \cmidrule{2-12}
& 5  & 0.612 & \heatcell{0.702} & 0.708 & 0.702 & \heatcell{0.705} & 0.542 & \heatcell{0.708} & 0.713 & 0.708 & \heatcell{0.710} \\ \cmidrule{2-12}
& 10 & 0.432 & \heatcell{0.765} & 0.770 & 0.765 & \heatcell{0.768} & 0.392 & \heatcell{0.789} & 0.794 & 0.789 & \heatcell{0.791} \\ \cmidrule{2-12}
& 15 & 0.312 & \heatcell{0.812} & 0.817 & 0.812 & \heatcell{0.814} & 0.281 & \heatcell{0.832} & 0.837 & 0.832 & \heatcell{0.834} \\ \midrule

\multirow{5}{*}{\textbf{DistilGPT2}} 
& 1  & 1.132 & \heatcell{0.431} & 0.438 & 0.431 & \heatcell{0.428} & 1.021 & \heatcell{0.449} & 0.455 & 0.449 & \heatcell{0.445} \\ \cmidrule{2-12}
& 2  & 1.021 & \heatcell{0.493} & 0.499 & 0.493 & \heatcell{0.490} & 0.926 & \heatcell{0.502} & 0.508 & 0.502 & \heatcell{0.499} \\ \cmidrule{2-12}
& 5  & 0.882 & \heatcell{0.542} & 0.548 & 0.542 & \heatcell{0.540} & 0.803 & \heatcell{0.551} & 0.557 & 0.551 & \heatcell{0.549} \\ \cmidrule{2-12}
& 10 & 0.745 & \heatcell{0.583} & 0.589 & 0.583 & \heatcell{0.581} & 0.673 & \heatcell{0.594} & 0.600 & 0.594 & \heatcell{0.592} \\ \cmidrule{2-12}
& 15 & 0.681 & \heatcell{0.615} & 0.621 & 0.615 & \heatcell{0.614} & 0.612 & \heatcell{0.624} & 0.630 & 0.624 & \heatcell{0.622} \\ \midrule

\multirow{5}{*}{\textbf{TinyT5}} 
& 1  & 1.224 & \heatcell{0.401} & 0.406 & 0.401 & \heatcell{0.399} & 1.118 & \heatcell{0.415} & 0.420 & 0.415 & \heatcell{0.412} \\ \cmidrule{2-12}
& 2  & 1.082 & \heatcell{0.445} & 0.451 & 0.445 & \heatcell{0.444} & 0.978 & \heatcell{0.456} & 0.461 & 0.456 & \heatcell{0.455} \\ \cmidrule{2-12}
& 5  & 0.932 & \heatcell{0.491} & 0.496 & 0.491 & \heatcell{0.490} & 0.841 & \heatcell{0.505} & 0.510 & 0.505 & \heatcell{0.504} \\ \cmidrule{2-12}
& 10 & 0.812 & \heatcell{0.531} & 0.536 & 0.531 & \heatcell{0.530} & 0.732 & \heatcell{0.543} & 0.548 & 0.543 & \heatcell{0.542} \\ \cmidrule{2-12}
& 15 & 0.723 & \heatcell{0.569} & 0.574 & 0.569 & \heatcell{0.568} & 0.654 & \heatcell{0.578} & 0.583 & 0.578 & \heatcell{0.577} \\ \midrule
\end{tabular}
}

 \vspace{-0.6em}

\resizebox{0.90\textwidth}{!}{%
\begin{tabular}{@{}|c|c|c|c|c|c|c|c|c|c|c|c|@{}}
\midrule
\multicolumn{2}{|c|}{} & \multicolumn{5}{c|}{\textbf{Round 2}} & \multicolumn{5}{c|}{\textbf{Round 3}} \\ \midrule
\textbf{Model} & \textbf{Epoch} & \textbf{Loss} & \textbf{Accuracy} & \textbf{Precision} & \textbf{Recall} & \textbf{F1-score} 
& \textbf{Loss} & \textbf{Accuracy} & \textbf{Precision} & \textbf{Recall} & \textbf{F1-score} \\ \midrule

\multirow{5}{*}{\textbf{DistilBERT}} 
& 1  & 0.879 & \heatcell{0.502} & 0.507 & 0.502 & \heatcell{0.498} & 0.841 & \heatcell{0.519} & 0.523 & 0.519 & \heatcell{0.516} \\ \cmidrule{2-12}
& 2  & 0.709 & \heatcell{0.611} & 0.616 & 0.611 & \heatcell{0.609} & 0.661 & \heatcell{0.627} & 0.631 & 0.627 & \heatcell{0.629} \\ \cmidrule{2-12}
& 5  & 0.498 & \heatcell{0.732} & 0.737 & 0.732 & \heatcell{0.734} & 0.457 & \heatcell{0.748} & 0.752 & 0.748 & \heatcell{0.750} \\ \cmidrule{2-12}
& 10 & 0.362 & \heatcell{0.803} & 0.808 & 0.803 & \heatcell{0.805} & 0.333 & \heatcell{0.818} & 0.822 & 0.818 & \heatcell{0.820} \\ \cmidrule{2-12}
& 15 & 0.258 & \heatcell{0.846} & 0.850 & 0.846 & \heatcell{0.848} & 0.235 & \heatcell{0.862} & 0.866 & 0.862 & \heatcell{0.864} \\ \midrule

\multirow{5}{*}{\textbf{DistilGPT2}} 
& 1  & 0.977 & \heatcell{0.472} & 0.478 & 0.472 & \heatcell{0.469} & 0.952 & \heatcell{0.481} & 0.486 & 0.481 & \heatcell{0.479} \\ \cmidrule{2-12}
& 2  & 0.883 & \heatcell{0.514} & 0.519 & 0.514 & \heatcell{0.511} & 0.855 & \heatcell{0.521} & 0.527 & 0.521 & \heatcell{0.519} \\ \cmidrule{2-12}
& 5  & 0.762 & \heatcell{0.562} & 0.568 & 0.562 & \heatcell{0.560} & 0.739 & \heatcell{0.569} & 0.574 & 0.569 & \heatcell{0.568} \\ \cmidrule{2-12}
& 10 & 0.638 & \heatcell{0.606} & 0.612 & 0.606 & \heatcell{0.604} & 0.614 & \heatcell{0.612} & 0.617 & 0.612 & \heatcell{0.611} \\ \cmidrule{2-12}
& 15 & 0.578 & \heatcell{0.635} & 0.641 & 0.635 & \heatcell{0.634} & 0.555 & \heatcell{0.642} & 0.648 & 0.642 & \heatcell{0.641} \\ \midrule

\multirow{5}{*}{\textbf{TinyT5}} 
& 1  & 1.052 & \heatcell{0.429} & 0.434 & 0.429 & \heatcell{0.427} & 1.011 & \heatcell{0.438} & 0.443 & 0.438 & \heatcell{0.437} \\ \cmidrule{2-12}
& 2  & 0.921 & \heatcell{0.468} & 0.473 & 0.468 & \heatcell{0.467} & 0.888 & \heatcell{0.474} & 0.479 & 0.474 & \heatcell{0.473} \\ \cmidrule{2-12}
& 5  & 0.793 & \heatcell{0.517} & 0.522 & 0.517 & \heatcell{0.516} & 0.762 & \heatcell{0.524} & 0.529 & 0.524 & \heatcell{0.523} \\ \cmidrule{2-12}
& 10 & 0.689 & \heatcell{0.554} & 0.559 & 0.554 & \heatcell{0.553} & 0.661 & \heatcell{0.562} & 0.567 & 0.562 & \heatcell{0.561} \\ \cmidrule{2-12}
& 15 & 0.612 & \heatcell{0.589} & 0.594 & 0.589 & \heatcell{0.588} & 0.586 & \heatcell{0.596} & 0.601 & 0.596 & \heatcell{0.595} \\ \midrule
\end{tabular}
}
\vspace{+0.2em}
\caption{Round improvement for all models without LoRA across different epochs.}
\label{tab:all-models-nolora}
\end{table*}

\begin{table*}[!htbp]
\centering

\resizebox{0.90\textwidth}{!}{%
\begin{tabular}{@{}|c|c|c|c|c|c|c|c|c|c|c|c|@{}}
\midrule
\multicolumn{2}{|c|}{} & \multicolumn{5}{c|}{\textbf{Round 0}} & \multicolumn{5}{c|}{\textbf{Round 1}} \\ \midrule
\textbf{Model} & \textbf{Epoch} & \textbf{Loss} & \textbf{Accuracy} & \textbf{Precision} & \textbf{Recall} & \textbf{F1-score}
& \textbf{Loss} & \textbf{Accuracy} & \textbf{Precision} & \textbf{Recall} & \textbf{F1-score} \\ \midrule

\multirow{5}{*}{\textbf{DistilBERT}}
& 1  & 0.937 & \heatcell{0.554} & 0.604 & 0.554 & \heatcell{0.564} & 0.787 & \heatcell{0.595} & 0.660 & 0.585 & \heatcell{0.511} \\ \cmidrule{2-12}
& 2  & 0.515 & \heatcell{0.879} & 0.884 & 0.879 & \heatcell{0.879} & 0.479 & \heatcell{0.924} & 0.791 & 0.733 & \heatcell{0.691} \\ \cmidrule{2-12}
& 5  & 0.312 & \heatcell{0.953} & 0.956 & 0.953 & \heatcell{0.953} & 0.294 & \heatcell{0.968} & 0.875 & 0.813 & \heatcell{0.789} \\ \cmidrule{2-12}
& 10 & 0.188 & \heatcell{0.984} & 0.985 & 0.984 & \heatcell{0.984} & 0.165 & \heatcell{0.988} & 0.977 & 0.975 & \heatcell{0.975} \\ \cmidrule{2-12}
& 15 & 0.121 & \heatcell{0.993} & 0.994 & 0.993 & \heatcell{0.993} & 0.106 & \heatcell{0.995} & 0.989 & 0.988 & \heatcell{0.989} \\ \midrule

\multirow{6}{*}{\textbf{DistilGPT2}}
& 1  & 1.021 & \heatcell{0.492} & 0.498 & 0.492 & \heatcell{0.488} & 0.888 & \heatcell{0.548} & 0.554 & 0.548 & \heatcell{0.544} \\ \cmidrule{2-12}
& 2  & 0.841 & \heatcell{0.682} & 0.688 & 0.682 & \heatcell{0.681} & 0.742 & \heatcell{0.698} & 0.704 & 0.698 & \heatcell{0.696} \\ \cmidrule{2-12}
& 3  & 0.721 & \heatcell{0.781} & 0.786 & 0.781 & \heatcell{0.780} & 0.631 & \heatcell{0.804} & 0.809 & 0.804 & \heatcell{0.802} \\ \cmidrule{2-12}
& 5  & 0.523 & \heatcell{0.854} & 0.858 & 0.854 & \heatcell{0.853} & 0.450 & \heatcell{0.865} & 0.869 & 0.865 & \heatcell{0.864} \\ \cmidrule{2-12}
& 10 & 0.351 & \heatcell{0.913} & 0.917 & 0.913 & \heatcell{0.912} & 0.299 & \heatcell{0.919} & 0.922 & 0.919 & \heatcell{0.918} \\ \cmidrule{2-12}
& 15 & 0.288 & \heatcell{0.936} & 0.940 & 0.936 & \heatcell{0.935} & 0.243 & \heatcell{0.942} & 0.945 & 0.942 & \heatcell{0.941} \\ \midrule

\multirow{5}{*}{\textbf{TinyT5}}
& 1  & 1.112 & \heatcell{0.442} & 0.450 & 0.442 & \heatcell{0.439} & 0.978 & \heatcell{0.465} & 0.471 & 0.465 & \heatcell{0.461} \\ \cmidrule{2-12}
& 2  & 0.945 & \heatcell{0.562} & 0.568 & 0.562 & \heatcell{0.560} & 0.841 & \heatcell{0.586} & 0.591 & 0.586 & \heatcell{0.584} \\ \cmidrule{2-12}
& 5  & 0.772 & \heatcell{0.649} & 0.655 & 0.649 & \heatcell{0.648} & 0.682 & \heatcell{0.673} & 0.679 & 0.673 & \heatcell{0.671} \\ \cmidrule{2-12}
& 10 & 0.623 & \heatcell{0.721} & 0.726 & 0.721 & \heatcell{0.720} & 0.542 & \heatcell{0.742} & 0.747 & 0.742 & \heatcell{0.741} \\ \cmidrule{2-12}
& 15 & 0.541 & \heatcell{0.768} & 0.773 & 0.768 & \heatcell{0.767} & 0.467 & \heatcell{0.789} & 0.794 & 0.789 & \heatcell{0.788} \\ \midrule
\end{tabular}
}

 \vspace{-0.6em}

\resizebox{0.90\textwidth}{!}{%
\begin{tabular}{@{}|c|c|c|c|c|c|c|c|c|c|c|c|@{}}
\midrule
\multicolumn{2}{|c|}{} & \multicolumn{5}{c|}{\textbf{Round 2}} & \multicolumn{5}{c|}{\textbf{Round 3}} \\ \midrule
\textbf{Model} & \textbf{Epoch} & \textbf{Loss} & \textbf{Accuracy} & \textbf{Precision} & \textbf{Recall} & \textbf{F1-score}
& \textbf{Loss} & \textbf{Accuracy} & \textbf{Precision} & \textbf{Recall} & \textbf{F1-score} \\ \midrule

\multirow{5}{*}{\textbf{DistilBERT}}
& 1  & 0.636 & \heatcell{0.717} & 0.460 & 0.465 & \heatcell{0.459} & 0.668 & \heatcell{0.675} & 0.718 & 0.675 & \heatcell{0.658} \\ \cmidrule{2-12}
& 2  & 0.442 & \heatcell{0.969} & 0.697 & 0.587 & \heatcell{0.503} & 0.623 & \heatcell{0.940} & 0.943 & 0.940 & \heatcell{0.939} \\ \cmidrule{2-12}
& 5  & 0.276 & \heatcell{0.982} & 0.794 & 0.672 & \heatcell{0.624} & 0.398 & \heatcell{0.987} & 0.989 & 0.987 & \heatcell{0.987} \\ \cmidrule{2-12}
& 10 & 0.142 & \heatcell{0.992} & 0.968 & 0.965 & \heatcell{0.966} & 0.228 & \heatcell{0.997} & 0.998 & 0.997 & \heatcell{0.997} \\ \cmidrule{2-12}
& 15 & 0.091 & \heatcell{0.997} & 0.984 & 0.983 & \heatcell{0.984} & 0.145 & \heatcell{0.999} & 0.999 & 0.999 & \heatcell{0.999} \\ \midrule

\multirow{6}{*}{\textbf{DistilGPT2}}
& 1  & 0.755 & \heatcell{0.603} & 0.609 & 0.603 & \heatcell{0.598} & 0.699 & \heatcell{0.633} & 0.639 & 0.633 & \heatcell{0.629} \\ \cmidrule{2-12}
& 2  & 0.642 & \heatcell{0.714} & 0.719 & 0.714 & \heatcell{0.710} & 0.581 & \heatcell{0.756} & 0.761 & 0.756 & \heatcell{0.754} \\ \cmidrule{2-12}
& 3  & 0.540 & \heatcell{0.826} & 0.830 & 0.826 & \heatcell{0.824} & 0.488 & \heatcell{0.849} & 0.854 & 0.849 & \heatcell{0.847} \\ \cmidrule{2-12}
& 5  & 0.377 & \heatcell{0.876} & 0.880 & 0.876 & \heatcell{0.875} & 0.342 & \heatcell{0.892} & 0.896 & 0.892 & \heatcell{0.891} \\ \cmidrule{2-12}
& 10 & 0.246 & \heatcell{0.924} & 0.927 & 0.924 & \heatcell{0.923} & 0.223 & \heatcell{0.936} & 0.940 & 0.936 & \heatcell{0.935} \\ \cmidrule{2-12}
& 15 & 0.198 & \heatcell{0.947} & 0.950 & 0.947 & \heatcell{0.947} & 0.176 & \heatcell{0.957} & 0.960 & 0.957 & \heatcell{0.957} \\ \midrule

\multirow{5}{*}{\textbf{TinyT5}}
& 1  & 0.902 & \heatcell{0.523} & 0.529 & 0.523 & \heatcell{0.519} & 0.856 & \heatcell{0.541} & 0.548 & 0.541 & \heatcell{0.538} \\ \cmidrule{2-12}
& 2  & 0.777 & \heatcell{0.612} & 0.618 & 0.612 & \heatcell{0.610} & 0.733 & \heatcell{0.629} & 0.635 & 0.629 & \heatcell{0.627} \\ \cmidrule{2-12}
& 5  & 0.611 & \heatcell{0.701} & 0.707 & 0.701 & \heatcell{0.699} & 0.571 & \heatcell{0.718} & 0.724 & 0.718 & \heatcell{0.717} \\ \cmidrule{2-12}
& 10 & 0.489 & \heatcell{0.764} & 0.769 & 0.764 & \heatcell{0.763} & 0.452 & \heatcell{0.781} & 0.786 & 0.781 & \heatcell{0.780} \\ \cmidrule{2-12}
& 15 & 0.423 & \heatcell{0.812} & 0.817 & 0.812 & \heatcell{0.811} & 0.388 & \heatcell{0.828} & 0.833 & 0.828 & \heatcell{0.827} \\ \midrule
\end{tabular}
}

\vspace{+0.2em}
\caption{Round improvement for all models with LoRA across different epochs.}
\label{tab:all-models-lora}
\end{table*}

\subsubsection{Dataset}
To ensure model compatibility with multi-label classification tasks, datasets undergo a preparation phase that involves loading data from CSV files, encoding target labels using Scikit-learn's LabelEncoder, creating multilabel targets by aggregating encoded label columns, and concatenating feature columns for model input. 
In particular, the textual labels contained in the target columns (i.e., \texttt{Attack\_type}, representing the category of attack, and \texttt{Attack\_label}, indicating its presence or absence) are converted into numerical values using Scikit-learn’s LabelEncoder.
Also, we performed a light feature preparation on the dataset.
The occurrences with null features were removed, resulting in 61 features for EDGE-IIoTset and 44 for TON-IoT. 
We learned in our previous work \cite{rondanini2025malware} how the best trade-off between performance result and computation and training time is obtained by performing the evaluation on a small subset (few-shot learning).
Thus, we decided to perform the evaluation test only on a few-shot learning setup.
For example, taking into account the publicly available EDGE-IIoTset dataset, we considered a number of samples sufficient to cover different categories of attacks, as depicted in Table \ref{tab:malwareDistribution}.
The number of selected samples was then divided into training and test sets with a split of 60\% and 40\%, respectively. 
To prevent mismatches, guarantee consistent numeric encoding, enabling cross-domain validation, and avoiding errors when reloading the model, the TON-IoT targets are mapped using the same names and encodings as EDGE-IIoTset. 

\subsubsection{Setting}
Our experiments were conducted on two different machines:  a high-performance system equipped with an AMD EPYC 7742 64-core Processor, 32GB RAM, and an NVIDIA A100-SXM4-80GB GPU, and 
a Google Colab free virtual machine featuring CPU 2 vCPU Intel Xeon (Broadwell/Skylake), up to 12.6 GB RAM, and an NVIDIA Tesla T4-16GB VRAM GPU.  
In the following, we do not specify the machine used in each experiment, as it only impacts the training time for fine-tuning, while the other performance indicators remain identical.


\subsection{Evaluation of Local Continuous Learning with LoRA}\label{subsec:local_learning}

The key architectural innovation introduced in this work is the integration of LoRA adapters to enable parameter-efficient continuous learning.
Their inclusion results in only a minor increase in model size, with measurements indicating an overhead of approximately 1.79 MB for DistilBERT, 1.77 MB for DistilGPT-2, and 0.59 MB for TinyT5.
Such additions represent less than 1\% of the full model footprint and are fully compatible with the memory budgets of edge-class hardware.

The first set of experiments focuses on the local adaptation capability of edge-deployed models.
To assess local adaptation capability, we take a pre-trained model (e.g., DistilBERT), perform a single fine-tuning round using locally available data, generate a LoRA module capturing the adaptations, and apply it back to the base model to emulate continuous single-device learning.
As summarized in Table~\ref{tab:all-models-nolora}, without LORA, traditional full fine-tuning led to gradual improvements in accuracy and F1-score across rounds, but remained inadequate for the target performance requirements.
Between Round~0 and Round~1, all models exhibited moderate learning progress, with DistilBERT improving from an initial F1-score of $0.814$ to $0.834$, DistilGPT-2 rising from $0.614$ to $0.622$, and TinyT5 increasing from $0.568$ to $0.577$ for 15 rounds.
Such results highlight the inefficiency of repeatedly retraining full models in constrained environments.

Then, by introducing LoRA adapters, we observed a significant improvement in both performance and stability, as reported in Table~\ref{tab:all-models-lora}.
By freezing the backbone and training only low-rank matrices, each model efficiently incorporated new malware knowledge across rounds while retaining prior representations.
To ensure comparability across rounds, predictions were computed from normalized logits (pre-softmax) rather than raw probabilities, avoiding inconsistencies caused by evolving class distributions.
From Round 1 of Table~\ref{tab:all-models-nolora} to Round 1 of Table~\ref{tab:all-models-lora}, all models demonstrated significant local improvement, DistilBERT rising from $0.834 $to $0.989$, DistilGPT-2 from $0.622$ to $0.941$, and TinyT5 from $0.577$ to $0.788$.
Comparing the two experimental settings, the benefits of integrating LoRA are evident across all models and evaluation rounds.
On average, LoRA-enhanced configurations achieved an improvement of approximately 25–35\% in F1-score and exhibited faster convergence, typically reaching near-optimal performance by the 5th to 10th epoch, compared to the slower and more gradual improvements observed in full fine-tuning.

\subsection{Evaluation of Global Continuous Learning}\label{subsec:decentralized_learning}
The second experiment evaluates the system’s ability to transfer knowledge across devices in a global setting.
Here, each edge node is initially fine-tuned on a distinct dataset (e.g., \textbf{Device 1} on Edge-IIoTset and \textbf{Device 2} on TON-IoT), thus learning only the malware patterns present in its own domain (see Figure \ref{fig:roundimprovement}).
After performing a single fine-tuning round using locally available data, we generate a LoRA module capturing the adaptations and apply it back to the base models.
In this way, the newly generated lora will cover both datasets.

From Round 2 of Table~\ref{tab:all-models-nolora} to Round 3 of Table~\ref{tab:all-models-lora}, the learning phase, where knowledge was exchanged among devices, improvements were observed across all models.
DistilBERT, for example, increased its F1-score from $0.864$ at Round~2 to $0.999$ at Round~3, while DistilGPT-2 improved from $0.641$ to $0.957$, and TinyT5 from $0.595$ to $0.827$.

In the baseline cross-dataset setting (without LoRA sharing), each model performed poorly when exposed to traffic from the other dataset, confirming the limited transferability of isolated fine-tuning.
However, when LoRA-based adapters trained on each device were exchanged and aggregated via the coordinating layer, both models showed substantial improvements.
After only one communication round, each device successfully identified previously unseen malware classes originating from the other dataset, achieving accuracy increases of up to 20–25\% compared to isolated training.

 \begin{figure}[!htbp]
\centering
 \includegraphics[width=\columnwidth, keepaspectratio]{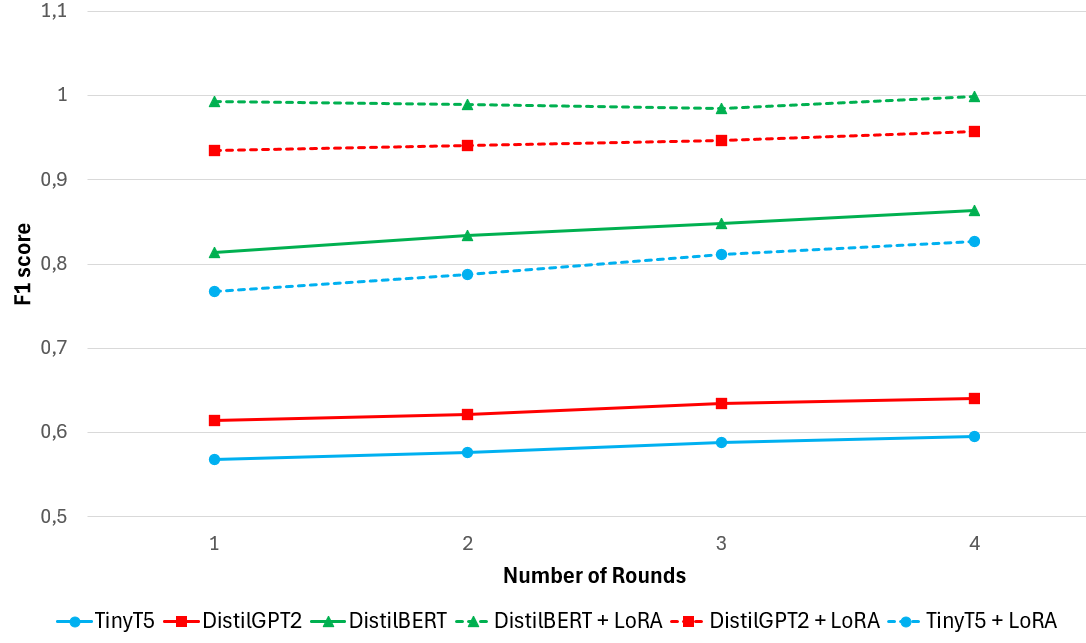}
 \caption{Progression of model performance across rounds}\label{fig:round-trends}
 \end{figure}

To better illustrate the progression of model performance across rounds, Figure~\ref{fig:round-trends} summarizes the evolution of F1-score for each model.
These findings show that LoRA-based training demonstrated superior stability across rounds, exhibiting reduced loss oscillations and minimal degradation when new data distributions were introduced. 
Moreover, LoRA provides a robust trade-off between computational efficiency and adaptive performance, making it highly suitable for incremental, resource-constrained edge devices.

\section{Related Work}\label{sec:applications}
 Malware detection using LLMs has been recently investigated, though not specifically targeting edge computing.   Most of the proposed solutions rely on BERT  and its variants. For example, SecureBERT \cite{aghaei2022securebert}, a model trained to automate malware and intrusion detection by interpreting cybersecurity texts, such as cyber threat intelligence documentation. Additionally, some proposals target Android malware detection,  like MalBERT  and MalBERTv2 \cite{rahali2021malbert_android,rahali2023malbertv2_android} that apply BERT for detecting Android malware via static source code analysis; 
\cite{ullah2022explainable}, that exploits  BERT-based transfer learning approach and a novel malware-to-image transformation for enhanced analysis;  AppPoet \cite{zhao2024apppoet} that leverages GPT-4 with multi-view prompt engineering to improve Android malware detection by incorporating detailed feature descriptions.
Windows malware detection through BERT-based dynamic analysis has been proposed in  \cite{xu2021malbert_windows}. \cite{sanchez2024transfer} presents an approach to malware detection with LLMs by examining the patterns of system calls generated by applications.  The work demonstrates the potential of LLMs in the context of malware detection through syscall analysis. More close to edge computing,   \cite{omar2024harnessing} presents a comparative analysis of LLMs, specifically BERT and GPT-2, for the detection of zero-day attacks on IoT devices.  The analysis results indicate that BERT outperforms GPT-2 across multiple metrics, demonstrating its effectiveness in generalizing to various attack types, but it does not provide results for the deployment of such models on the device.

Additionally,  the integration of large language models into edge computing has recently raised attention in the research community, and in particular, the paradigm of continuous learning or incremental learning. Several recent studies have examined the challenges and methodologies of these approaches, proposing detailed surveys.
 
As an example,  \cite{jovanovic2025towards} provides a detailed discussion of incremental learning paradigms for LLMs, including continual learning, meta-learning, and mixture-of-experts, by identifying the challenges, such as catastrophic forgetting and scalability. However, the analysis remains theoretical, offering little guidance for real-world deployment in settings with limited resources like edge devices. Another point of view is presented in \cite{guo2025comprehensive}, where approaches for continual learning techniques are categorized into architecture-based, regularization-based, and replay-based. While this perspective highlights the adaptability of generative models to evolving data, it is less relevant to classification tasks such as malware detection. Also, generative replay often implies high resource consumption, raising concerns about scalability in constrained settings.
 
Instead, \cite{zheng2025towards} analyzes the strategies for lifelong learning by emphasizing the integration of external knowledge sources and the potential of model expansion techniques. Its key contribution is the recognition of dynamic environments where models must balance plasticity and stability. However, it relies heavily on external retrieval or architectural growth, which can be impractical for edge deployment, where storage and computation are limited.

Detailed classifications are also provided in other works.
For example, \cite{wu2301continual} introduces a structured view of continual learning methods by distinguishing between continual pretraining, instruction tuning, and alignment.
Its contribution clarifies how adaptation can be staged throughout the model lifecycle. However, the survey focuses on general-purpose NLP tasks, without addressing constraints or requirements specific to security applications such as malware detection. An alternative classification including approaches such as continual pretraining, domain-adaptive pretraining, and continual fine-tuning is proposed in \cite{shi2024continual}. Despite addressing evaluation protocols and focusing on methodological rigor and evaluation metrics, the reliance on large-scale pretraining and domain-specific fine-tuning makes many reviewed techniques difficult to apply in edge scenarios, where retraining from scratch is not feasible.

Other contributions have focused on LoRA and its variants as parameter-efficient solutions for continual learning. Notable examples to mitigate catastrophic forgetting are \cite{wang2023orthogonal}  and \cite{liang2025gated}, which exploit orthogonality constraints in low-rank adaptation and expand LoRA branches for each task and combine them through gating. However,  the first approach adds mathematical and computational complexity, potentially limiting its feasibility in low-resource contexts. Instead, the latter approach enhances flexibility and accuracy, but the progressive accumulation of adapters can lead to inefficiencies when the number of tasks grows.
 Other works focus on improving the trade-off between stability and plasticity (i.e., the model's ability to learn and adapt to new tasks or information), such as  \cite{wu2025sd}. The empirical performance in class-incremental learning benchmarks is remarkable, but it does not explicitly address deployment in resource-limited scenarios, which remain a primary concern in edge computing. An ensemble strategy in which a sequence of LoRAs is learned and combined through attention is proposed in \cite{liu2024learning}.
This method effectively captures task diversity but introduces additional inference overhead due to the attention mechanism, which may lower performance on lightweight devices.

Other approaches explore hybrid strategies through the combination of different approaches. As an example, \cite{borhanifard2025combining} demonstrates that the combination of replay buffers and LoRA adapters provides higher retention of prior knowledge.
Moreover, \cite{wistuba2023continual} proposes a combination of memory replay with LoRA in order to mitigate catastrophic forgetting. However, these methods have limitations, such as inheriting replay limitations, namely storage overhead and privacy risks, which are relevant in security-sensitive environments and focus only on general NLP tasks.

In contrast to these studies, our work explicitly targets the problem of malware detection at the edge, a domain characterized by severe resource constraints and rapidly evolving threat landscapes. 
By adopting small language models combined with LoRA, we demonstrate that high accuracy and adaptability can be achieved without incurring the computational cost of large-scale pretraining or the storage overhead of replay. 
Moreover, unlike prior proposals that emphasize general-purpose NLP or generative tasks, our architecture is tailored to classification in security-critical environments, with specific design choices such as logit-based prediction aggregation to ensure stability across incremental rounds.

 \section{Conclusions}\label{sec:conclusions}

This work explored the integration of parameter-efficient fine-tuning through Low-Rank Adaptation (LoRA) into transformer-based models deployed within a distributed edge computing environment for continuous malware detection.
Through comparative experiments involving DistilBERT, DistilGPT-2, and TinyT5, both with and without LoRA, we demonstrated that LoRA enables models to maintain high predictive accuracy while dramatically reducing the computational and memory requirements associated with full fine-tuning. 
The results consistently show that models equipped with LoRA converge faster and adapt more effectively to novel data patterns across consecutive learning rounds, confirming its robustness for incremental and on-device learning scenarios. Furthermore, LoRA mitigates catastrophic forgetting by preserving the core model parameters and confining updates to low-rank adapters, ensuring efficient knowledge integration without compromising previously learned representations. 
These findings establish LoRA as a practical and scalable solution for maintaining intelligent edge systems capable of autonomous, continuous adaptation.
Future research will focus on exploring cross-device synchronization strategies to further enhance the resilience of edge-based AI infrastructures.






\section*{Acknowledgment}
 This research work has been partially supported by Cisco Research under the research grant award:  ''Malware Detection for Edge-based Computing'' (Id 88749009) and by the  SERICS project (PE00000014) under the NRRP MUR program funded by the EU - NGEU.
 Views and opinions expressed are, however, those of the authors only and do not necessarily reflect those of the European Union or the Italian MUR. Neither the European Union nor the Italian MUR can be held responsible for them.

\bibliographystyle{IEEEtran}
\bibliography{biblio}

@ARTICLE{Ferrag_EdgeIOT,
  author={Ferrag, Mohamed Amine and Friha, Othmane and Hamouda, Djallel and Maglaras, Leandros and Janicke, Helge},
  journal={IEEE Access}, 
  title={Edge-IIoTset: A New Comprehensive Realistic Cyber Security Dataset of IoT and IIoT Applications for Centralized and Federated Learning}, 
  year={2022},
  volume={10},
  number={},
  pages={40281-40306},
  doi={10.1109/ACCESS.2022.3165809}}

@article{moustafa2021new,
  title={A new distributed architecture for evaluating AI-based security systems at the edge: Network TON\_IoT datasets},
  author={Moustafa, Nour},
  journal={Sustainable Cities and Society},
  volume={72},
  pages={},
  year={2021},
  publisher={Elsevier}
}

@article{xu2024survey,
  title={A survey on knowledge distillation of large language models},
  author={Xu, Xiaohan and Li, Ming and Tao, Chongyang and Shen, Tao and Cheng, Reynold and Li, Jinyang and Xu, Can and Tao, Dacheng and Zhou, Tianyi},
  journal={arXiv preprint arXiv:2402.13116},
  year={2024}
}

@inproceedings{rondanini2024large,
  title={Large Language Models to Enhance Malware Detection in Edge Computing},
  author={Rondanini, Christian and Carminati, Barbara and Ferrari, Elena and Kundu, Ashish and Jajoo, Akshay},
  booktitle={2024 IEEE 6th International Conference on Trust, Privacy and Security in Intelligent Systems, and Applications (TPS-ISA)},
  year={2024}
}

@inproceedings{omar2024harnessing,
  title={Harnessing LLMs for IoT Malware Detection: A Comparative Analysis of BERT and GPT-2},
  author={Omar, Marwan and Zangana, Hewa Majeed and Al-Karaki, Jamal N and Mohammed, Derek},
  booktitle={2024 8th International Symposium on Multidisciplinary Studies and Innovative Technologies (ISMSIT)},
  pages={1--6},
  year={2024},
  organization={IEEE}
}

@article{sanchez2024transfer,
  title={Transfer Learning in Pre-Trained Large Language Models for Malware Detection Based on System Calls},
  author={S{\'a}nchez, Pedro Miguel S{\'a}nchez and Celdr{\'a}n, Alberto Huertas and Bovet, G{\'e}r{\^o}me and P{\'e}rez, Gregorio Mart{\'\i}nez},
  journal={arXiv preprint arXiv:2405.09318},
  year={2024}
}

@inproceedings{rahali2021malbert_android,
  title={Malbert: Malware detection using bidirectional encoder representations from transformers},
  author={Rahali, Abir and Akhloufi, Moulay A},
  booktitle={2021 IEEE International Conference on Systems, Man, and Cybernetics (SMC)},
  pages={},
  year={2021},
  organization={IEEE}
}

@article{rahali2023malbertv2_android,
  title={Malbertv2: Code aware bert-based model for malware identification},
  author={{Rahali, Abir and Akhloufi, Moulay A}},
  journal={Big Data and Cognitive Computing},
  volume={},
  number={},
  pages={},
  year={2023},
  publisher={MDPI}
}

@article{xu2021malbert_windows,
  title={Malbert: A novel pre-training method for malware detection},
  author={Xu, Zhifeng and Fang, Xianjin and Yang, Gaoming},
  journal={Computers \& Security},
  volume={},
  pages={},
  year={2021},
  publisher={Elsevier}
}

@article{devlin2018bert,
  title={Bert: Pre-training of deep bidirectional transformers for language understanding},
  author={Devlin, Jacob and Chang, Ming-Wei and Lee, Kenton and Toutanova, Kristina},
  journal={arXiv preprint arXiv:1810.04805},
  year={2018}
}

@article{sanh2019distilbert,
  title={DistilBERT, a distilled version of BERT: smaller, faster, cheaper and lighter},
  author={Sanh, Victor and Debut, Lysandre and Chaumond, Julien and Wolf, Thomas},
  journal={arXiv preprint arXiv:1910.01108},
  year={2019}
}

@article{gptcore,
  title={Improving language understanding by generative pre-training},
  author={Radford, Alec and Narasimhan, Karthik and Salimans, Tim and Sutskever, Ilya and others},
  year={2018},
  journal={Hayate-lab},
  publisher={San Francisco, CA, USA}
}

@inproceedings{aghaei2022securebert,
  title={Securebert: A domain-specific language model for cybersecurity},
  author={Aghaei, Ehsan and Niu, Xi and Shadid, Waseem and Al-Shaer, Ehab},
  booktitle={International Conference on Security and Privacy in Communication Systems},
  pages={}, 
  year={2022},
  organization={Springer}
}

@article{zhao2024apppoet,
  title={AppPoet: Large Language Model based Android malware detection via multi-view prompt engineering},
  author={Zhao, Wenxiang and Wu, Juntao and Meng, Zhaoyi},
  journal={arXiv preprint arXiv:2404.18816},
  volume={},
  pages={},
  year={2024}
}

@article{ullah2022explainable,
  title={Explainable malware detection system using transformers-based transfer learning and multi-model visual representation},
  author={Ullah, Farhan and Alsirhani, Amjad and Alshahrani, Mohammed Mujib and Alomari, Abdullah and Naeem, Hamad and Shah, Syed Aziz},
  journal={Sensors},
  volume={},
  number={},
  pages={},
  year={2022},
  publisher={MDPI}
}

@article{raffel2020exploring,
  title={Exploring the limits of transfer learning with a unified text-to-text transformer},
  author={Raffel, Colin and Shazeer, Noam and Roberts, Adam and Lee, Katherine and Narang, Sharan and Matena, Michael and Zhou, Yanqi and Li, Wei and Liu, Peter J},
  journal={Journal of machine learning research},
  volume={},
  number={},
  pages={},
  year={2020}
}

@inproceedings{panebianco2025guessing,
  author    = {Francesco Panebianco and Andrea Isgr{\`o} and Stefano Longari and Stefano Zanero and Michele Carminati},
  title     = {Guessing As A Service: Large Language Models Are Not Yet Ready For Vulnerability Detection},
  booktitle = {Proceedings of ITASEC25 (Joint National Conference on Cybersecurity (ITASEC \& SERICS))},
  year      = {2025}
}

@article{brown1992class,
  title={Class-based n-gram models of natural language},
  author={Brown, Peter F and Della Pietra, Vincent J and Desouza, Peter V and Lai, Jennifer C and Mercer, Robert L},
  journal={Computational linguistics},
  volume={},
  number={},
  pages={},
  year={1992}
}

@inproceedings{mikolov2011strategies,
  title={Strategies for training large scale neural network language models},
  author={Mikolov, Tom{\'a}{\v{s}} and Deoras, Anoop and Povey, Daniel and Burget, Luk{\'a}{\v{s}} and {\v{C}}ernock{\`y}, Jan},
  booktitle={2011 IEEE Workshop on Automatic Speech Recognition \& Understanding},
  pages={},
  year={2011},
  organization={}
}

@article{sutskever2014sequence,
  title={Sequence to sequence learning with neural networks},
  author={Sutskever, Ilya and Vinyals, Oriol and Le, Quoc V},
  journal={Advances in neural information processing systems},
  volume={},
  year={2014}
}

@article{jovanovic2025towards,
  title={Towards Incremental Learning in Large Language Models: A Critical Review},
  author={Jovanovic, Mladjan and Voss, Peter},
  journal={Expert Systems},
  volume={42},
  number={10},
  pages={e70127},
  year={2025},
  publisher={Wiley Online Library}
}

@article{zheng2025towards,
  title={Towards lifelong learning of large language models: A survey},
  author={Zheng, Junhao and Qiu, Shengjie and Shi, Chengming and Ma, Qianli},
  journal={ACM Computing Surveys},
  volume={57},
  number={8},
  pages={1--35},
  year={2025},
  publisher={ACM New York, NY}
}

@article{shi2024continual,
  title={Continual learning of large language models: A comprehensive survey},
  author={Shi, Haizhou and Xu, Zihao and Wang, Hengyi and Qin, Weiyi and Wang, Wenyuan and Wang, Yibin and Wang, Zifeng and Ebrahimi, Sayna and Wang, Hao},
  journal={ACM Computing Surveys},
  year={2024},
  publisher={ACM New York, NY}
}

@article{xiao2019edge,
  title={Edge computing security: State of the art and challenges},
  author={Xiao, Yinhao and Jia, Yizhen and Liu, Chunchi and Cheng, Xiuzhen and Yu, Jiguo and Lv, Weifeng},
  journal={Proceedings of the IEEE},
  volume={107},
  number={8},
  pages={1608--1631},
  year={2019},
  publisher={IEEE}
}

@article{ferdous2025survey,
  title={A Survey on ML Techniques for Multi-Platform Malware Detection: Securing PC, Mobile Devices, IoT, and Cloud Environments},
  author={Ferdous, Jannatul and Islam, Rafiqul and Mahboubi, Arash and Islam, Md Zahidul},
  journal={Sensors (Basel, Switzerland)},
  volume={25},
  number={4},
  pages={1153},
  year={2025}
}

@article{han2024parameter,
  title={Parameter-efficient fine-tuning for large models: A comprehensive survey},
  author={Han, Zeyu and Gao, Chao and Liu, Jinyang and Zhang, Jeff and Zhang, Sai Qian},
  journal={arXiv preprint arXiv:2403.14608},
  year={2024}
}

@article{beltran2023decentralized,
 title={Decentralized federated learning: Fundamentals, state of the art, frameworks, trends, and challenges},
 author={Beltr{\'a}n, Enrique Tom{\'a}s Mart{\'\i}nez and P{\'e}rez, Mario Quiles and S{\'a}nchez, Pedro Miguel S{\'a}nchez and Bernal, Sergio L{\'o}pez and Bovet, G{\'e}r{\^o}me and P{\'e}rez, Manuel Gil and P{\'e}rez, Gregorio Mart{\'\i}nez and Celdr{\'a}n, Alberto Huertas},
 journal={IEEE Communications Surveys \& Tutorials},
 volume={25},
 number={4},
 pages={2983--3013},
 year={2023},
 publisher={IEEE}
}

@article{guo2025comprehensive,
  title={A Comprehensive Survey on Continual Learning in Generative Models},
  author={Guo, Haiyang and Zeng, Fanhu and Zhu, Fei and Wang, Jiayi and Wang, Xukai and Zhou, Jingang and Zhao, Hongbo and Liu, Wenzhuo and Ma, Shijie and Zhang, Xu-Yao and others},
  journal={arXiv preprint arXiv:2506.13045},
  year={2025}
}

@article{wang2023orthogonal,
  title={Orthogonal subspace learning for language model continual learning},
  author={Wang, Xiao and Chen, Tianze and Ge, Qiming and Xia, Han and Bao, Rong and Zheng, Rui and Zhang, Qi and Gui, Tao and Huang, Xuanjing},
  journal={arXiv preprint arXiv:2310.14152},
  year={2023}
}

@article{liang2025gated,
  title={Gated Integration of Low-Rank Adaptation for Continual Learning of Language Models},
  author={Liang, Yan-Shuo and Li, Wu-Jun},
  journal={arXiv preprint arXiv:2505.15424},
  year={2025}
}

@article{wu2025sd,
  title={Sd-lora: Scalable decoupled low-rank adaptation for class incremental learning},
  author={Wu, Yichen and Piao, Hongming and Huang, Long-Kai and Wang, Renzhen and Li, Wanhua and Pfister, Hanspeter and Meng, Deyu and Ma, Kede and Wei, Ying},
  journal={arXiv preprint arXiv:2501.13198},
  year={2025}
}

@article{liu2024learning,
  title={Learning attentional mixture of loras for language model continual learning},
  author={Liu, Jialin and Wu, Jianhua and Liu, Jie and Duan, Yutai},
  journal={arXiv preprint arXiv:2409.19611},
  year={2024}
}

@article{borhanifard2025combining,
  title={Combining replay and LoRA for continual learning in natural language understanding},
  author={Borhanifard, Zeinab and Faili, Heshaam and Yaghoobzadeh, Yadollah},
  journal={Computer Speech \& Language},
  volume={90},
  pages={101737},
  year={2025},
  publisher={Elsevier}
}

@article{wistuba2023continual,
  title={Continual learning with low rank adaptation},
  author={Wistuba, Martin and Sivaprasad, Prabhu Teja and Balles, Lukas and Zappella, Giovanni},
  journal={arXiv preprint arXiv:2311.17601},
  year={2023}
}

@article{wu2301continual,
  title={Continual learning for large language models: A survey},
  author={Wu, T and Luo, L and Li, YF and Pan, S and Vu, TT and Haffari, G},
  journal={arXiv preprint arXiv:2301.07082},
  year={2024}
}

@article{rondanini2025malware,
  title={Malware Detection at the Edge with Lightweight LLMs: A Performance Evaluation},
  author={Rondanini, Christian and Carminati, Barbara and Ferrari, Elena and Kundu, Ashish and Gaudiano, Antonio},
  journal={ACM Transactions on Internet Technology},
  year={2025},
  publisher={ACM New York, NY}
}

@inproceedings{yu2024boosting,
  title={Boosting continual learning of vision-language models via mixture-of-experts adapters},
  author={Yu, Jiazuo and Zhuge, Yunzhi and Zhang, Lu and Hu, Ping and Wang, Dong and Lu, Huchuan and He, You},
  booktitle={Proceedings of the IEEE/CVF Conference on Computer Vision and Pattern Recognition},
  pages={23219--23230},
  year={2024}
}

@article{li2017learning,
  title={Learning without forgetting},
  author={Li, Zhizhong and Hoiem, Derek},
  journal={IEEE transactions on pattern analysis and machine intelligence},
  volume={40},
  number={12},
  pages={2935--2947},
  year={2017},
  publisher={IEEE}
}

@inproceedings{zenke2017continual,
  title={Continual learning through synaptic intelligence},
  author={Zenke, Friedemann and Poole, Ben and Ganguli, Surya},
  booktitle={International conference on machine learning},
  pages={3987--3995},
  year={2017},
  organization={PMLR}
}

@article{rolnick2019experience,
  title={Experience replay for continual learning},
  author={Rolnick, David and Ahuja, Arun and Schwarz, Jonathan and Lillicrap, Timothy and Wayne, Gregory},
  journal={Advances in neural information processing systems},
  volume={32},
  year={2019}
}

@article{shin2017continual,
  title={Continual learning with deep generative replay},
  author={Shin, Hanul and Lee, Jung Kwon and Kim, Jaehong and Kim, Jiwon},
  journal={Advances in neural information processing systems},
  volume={30},
  year={2017}
}

@article{lopez2017gradient,
  title={Gradient episodic memory for continual learning},
  author={Lopez-Paz, David and Ranzato, Marc'Aurelio},
  journal={Advances in neural information processing systems},
  volume={30},
  year={2017}
}

@article{xiang2023language,
  title={Language models meet world models: Embodied experiences enhance language models},
  author={Xiang, Jiannan and Tao, Tianhua and Gu, Yi and Shu, Tianmin and Wang, Zirui and Yang, Zichao and Hu, Zhiting},
  journal={Advances in neural information processing systems},
  volume={36},
  pages={75392--75412},
  year={2023}
}

@article{hu2022lora,
  title={Lora: Low-rank adaptation of large language models.},
  author={Hu, Edward J and Shen, Yelong and Wallis, Phillip and Allen-Zhu, Zeyuan and Li, Yuanzhi and Wang, Shean and Wang, Lu and Chen, Weizhu and others},
  journal={ICLR},
  volume={1},
  number={2},
  pages={3},
  year={2022}
}

@article{dettmers2023qlora,
  title={Qlora: Efficient finetuning of quantized llms},
  author={Dettmers, Tim and Pagnoni, Artidoro and Holtzman, Ari and Zettlemoyer, Luke},
  journal={Advances in neural information processing systems},
  volume={36},
  pages={10088--10115},
  year={2023}
}

@article{li2021prefix,
  title={Prefix-tuning: Optimizing continuous prompts for generation},
  author={Li, Xiang Lisa and Liang, Percy},
  journal={arXiv preprint arXiv:2101.00190},
  year={2021}
}

@article{zaken2021bitfit,
  title={Bitfit: Simple parameter-efficient fine-tuning for transformer-based masked language-models},
  author={Zaken, Elad Ben and Ravfogel, Shauli and Goldberg, Yoav},
  journal={arXiv preprint arXiv:2106.10199},
  year={2021}
}

@inproceedings{Lester2021ThePO,
  title={The Power of Scale for Parameter-Efficient Prompt Tuning},
  author={Brian Lester and Rami Al-Rfou and Noah Constant},
  booktitle={Conference on Empirical Methods in Natural Language Processing},
  year={2021},
  url={https://api.semanticscholar.org/CorpusID:233296808}
}

@article{distilbert,
  title={DistilBERT, a distilled version of BERT: smaller, faster, cheaper and lighter},
  author={Sanh, Victor and Debut, Lysandre and Chaumond, Julien and Wolf, Thomas},
  journal={arXiv preprint arXiv:1910.01108},
  year={2019}
}

@misc{distilgpt2,
  author       = {Hugging Face},
  title        = {DistilGPT2},
  year         = 2020,
  note         = {\url{https://huggingface.co/distilgpt2}}
}

@article{vaswani2017attention,
  title={Attention is all you need},
  author={Vaswani, Ashish and Shazeer, Noam and Parmar, Niki and Uszkoreit, Jakob and Jones, Llion and Gomez, Aidan N and Kaiser, {\L}ukasz and Polosukhin, Illia},
  journal={Advances in neural information processing systems},
  volume={30},
  year={2017}
}

\end{document}